\documentclass[preprint,12pt]{elsarticle}

\usepackage{amssymb}
\usepackage{mathrsfs}
\usepackage{mathrsfs}
\usepackage{mathrsfs}
\usepackage{amsfonts}
\usepackage{amsfonts}
\usepackage{amsfonts}
\usepackage{mathrsfs}
\usepackage{amsmath}
\usepackage{graphicx}
\usepackage{multirow}
\usepackage{caption2}
\usepackage{float}
\usepackage{booktabs}
\usepackage{array}

\numberwithin{equation}{section}

\journal{Signal Processing}
\begin{document}
\begin{frontmatter}



\title{Sparse Solution of Underdetermined Linear Equations via Adaptively Iterative Thresholding}


\author[1]{Jinshan Zeng\footnote{ E-mail: jsh.zeng@gmail.com}}
\author[1]{Shaobo Lin\footnote{ E-mail: sblin1983@gmail.com}}
\author[1]{Zongben Xu\footnote{Corresponding author,E-mail: zbxu@mail.xjtu.edu.cn}}

\address[1]{
School of Mathematics and Statistics,
Xi'an Jiaotong University, Xi'an, 710049, China}

\begin{abstract}
Finding the sparset solution of an underdetermined system of linear equations $y=Ax$
has attracted considerable attention in recent years.
Among a large number of algorithms,
iterative thresholding algorithms are recognized as one of the most efficient and important classes of algorithms.
This is mainly due to their low computational complexities, especially for large scale applications.
The aim of this paper is to provide guarantees on the global convergence of a wide class of iterative thresholding algorithms.
Since the thresholds of the considered algorithms are set adaptively at each iteration,
we call them adaptively iterative thresholding (AIT) algorithms.
As the main result, we show that as long as $A$ satisfies a certain coherence property,
AIT algorithms can find the correct support set within finite iterations,
and then converge to the original sparse solution exponentially fast once the correct support set has been identified.
Meanwhile, we also demonstrate that AIT algorithms are robust to the algorithmic parameters.
In addition, it should be pointed out that most of the existing iterative thresholding algorithms such as hard, soft, half
and smoothly clipped absolute deviation (SCAD) algorithms are included in the class of AIT algorithms studied in this paper.
\end{abstract}

\end{frontmatter}
\section{Introduction}

Finding the sparsest solution of an undertermined system of linear equations
is an important problem emerged in many applications (especially, in compressed sensing {\cite{Donoho06}}, {\cite{Candes06}}).
Generally, an undertermined system of linear equations can be described as
\begin{equation}
y=Ax,
\label{Equation}
\end{equation}
where $y\in \mathbf{R}^M$ and $A\in \mathbf{R}^{M\times N}$ ($M<N$) are known,
$x=(x_1,\dots,x_N)^T\in \mathbf{R}^N$ is unknown.
Thus, finding the sparsest solution of the equations (\ref{Equation}) can be mathematically modeled as the following $l_0$-minimization, that is,
\begin{equation}
\min_{x\in \mathbf{R}^N} {\|x\|_0} \ \ \text{s.t.}\ y=A x,
\label{L0MinExact}
\end{equation}
where $\|x\|_0$ denotes the number of the nonzero components of $x$ and is formally called the $l_0$-norm.
However, the problem (\ref{L0MinExact}) is NP-hard and generally intractable for computing.

Instead, there are mainly two classes of methods, that is, the greedy and relaxed methods for approximately solving the problem (\ref{L0MinExact}).
The basic idea of the greedy method is that a sparse solution is refined iteratively by successively identifying one or more components
that yield the greatest improvement in quality {\cite{MallatMP1993}}.
There are many commonly used greedy algorithms such as orthogonal matching pursuit (OMP) {\cite{PatiOMP1993}}, {\cite{TroppOMP2007}}, stagewise OMP (StOMP) {\cite{DonohoStOMP}}, regularized OMP (ROMP) {\cite{Needell2010ROMP}}, compressive sampling matching pursuit (CoSaMP) {\cite{CoSaMP}} and subspace pursuit {\cite{DaiSP}}.
The greedy algorithms can be quite fast, especially in the ultra-sparse case, and also may be very efficient at certain situations such as the dictionary contains a continuum of elements {\cite{TroppGreedyAlg}}.
However, the performance of the greedy algorithms can not be guaranteed when the signal is not very sparse or the level of the observational noise is relatively high.

The relaxed method converts the combinatorial $l_0$-minimization into a more tractable model via replacing the $l_0$-norm with a certain nonnegative and continuous function $P(\cdot)$, that is,
\begin{equation}
\min_{x\in \mathbf{R}^N} {P(x)} \ \ \text{s.t.}\ y=A x.
\label{PMinExact}
\end{equation}
One of the most important cases is the $l_1$-minimization (also known as \textit{Basis Pursuit} (BP) {\cite{Donoho98}}) with $P(x) = \|x\|_1$,
where $\|x\|_1 = \sum_{i=1}^N |x_i|$ is called the $l_1$-norm.
The $l_1$-minimization is a convex optimization problem and thus can be efficiently solved.
Because of this, the $l_1$-minimization gets its popularity and has been accepted as a very useful tool for solution to sparsity problems.
Nevertheless, it cannot promote further sparsity when applied to compressed sensing
{\cite{Chartrand2007}}, {\cite{Chartrand2008}}, {\cite{XuHalf2012}}, {\cite{L1/2TNN}}, {\cite{Candes2008RL1}}.
Moreover, many nonconvex functions were proposed as relaxations of the $l_0$-norm.
Some typical nonconvex examples are the $l_q$-norm ($0<q<1$) {\cite{Chartrand2007}}, {\cite{Chartrand2008}}, {\cite{XuHalf2012}}, {\cite{L1/2TNN}}, smoothly clipped absolute deviation (SCAD) {\cite{FanSCAD}} and minimax concave penalty (MCP) {\cite{ZhangMCP2010}}.
As compared with the $l_1$-minimization, the nonconvex relaxed models can usually induce better sparsity,
however, they are generally more difficult to be solved.

There are mainly two kinds of algorithms to solve the constrained optimization problem (\ref{PMinExact}).
The first one is the iteratively reweighted algorithm.
Two of the most important iteratively reweighted algorithms are the reweighted $l_1$-minimization {\cite{Candes2008RL1}} and iteratively reweighted least squares (IRLS) {\cite{FOCUSS}}, {\cite{DaubechiesIRLS}} algorithms.
One of the main advantages of this kind of algorithms is that they can be used to solve a general model (\ref{PMinExact}).
However, the computational complexities of these algorithms are usually relatively high.
The other one is commonly called the regularization method, which transforms the constrained optimization problem (\ref{PMinExact}) into the following unconstrained optimization problem via introducing a regularization parameter
\begin{equation}
\min_{x\in \mathbf{R}^N} \{\|Ax - y\|_2^2+\lambda P(x)\},
\label{PRegProb}
\end{equation}
where $\lambda>0$ is a regularization parameter.
There are many algorithms for solving the regularization model (\ref{PRegProb}).
Particularly, for some special $P(x)$ such as the $l_0$-norm, $l_q$-norms ($q=1,2/3,1/2$), SCAD and MCP,
the regularization models (\ref{PRegProb}) can permit the thresholding representations and thus
yield the corresponding iterative thresholding algorithms {\cite{L1/2TNN}}, {\cite{DaubechiesSoft04}}, {\cite{L2/3Cao2013}}, {\cite{BlumensathHard08}}.
Intuitively, an iterative thresholding algorithm can be seen as a procedure of
Landweber iteration projected by a certain thresholding operator.
Compared to the aforementioned algorithms including the greedy, BP and iteratively reweighted algorithms,
iterative thresholding algorithms can be implemented fast and
have almost the least computational complexity for large scale problems {\cite{Qian20011}}, {\cite{Zeng20012SAR}}, {\cite{Zeng20013AccSAR}}.
So far, most of theoretical results of the iterative thresholding algorithms were developed for the regularization model (\ref{PRegProb}) with fixed $\lambda$.
However, it is in general difficult to determine an appropriate regularization parameter $\lambda$, especially when $P$ is nonconvex.

Alternatively, some adaptive strategies for setting the regularization parameters were proposed for iterative thresholding algorithms.
One of the commonly used strategies is to set the regularization parameter adaptively according to a specified sparsity level at each iteration.
Once the specified sparsity level is given, the number of nonzero components of vector at each iteration is also determined.
In practice, the specified sparsity level is desired to be a good estimation of the true sparsity level.
This strategy was first adopted to the iterative hard thresholding algorithm (called hard algorithm for short) in {\cite{Blumensath08CS}},
and later the iterative soft {\cite{MalekiITA2009}} (called soft algorithm for short) and half {\cite{L1/2TNN}} (called half algorithm for short) thresholding algorithms.
The convergence of  hard algorithm was justified when $A$ satisfies a certain restricted isometry property (RIP) {\cite{Blumensath08CS}}.
Later, Maleki investigated the convergence of both hard and soft algorithms in terms of the coherence {\cite{MalekiITA2009}}.
Both in the analysis of {\cite{Blumensath08CS}} and {\cite{MalekiITA2009}},
the specified sparsity levels of AIT algorithms are set to be the true sparsity level of the original sparse solution,
however, which is commonly unknown in practice.
Therefore, the robustness of AIT algorithms to the specified sparsity levels is very important in practice and worth of investigation.
Moreover, besides the hard and soft algorithms,
there are many other AIT algorithms such as half, SCAD, MCP algorithms which are widely used in signal processing, variable selection and feature extraction.
However, as far as we know, there are lack of the corresponding theoretical guarantees on the global convergence of these algorithms for sparse solution to the underdetermined linear equations.
Thus, the theoretical performance of these AIT algorithms should be further studied.

In this paper,
we consider the global convergence a wide class of adaptively iterative thresholding (AIT) algorithms for sparse solution to an underdetermined system of linear equations.
The associated thresholding functions satisfy some basic assumptions including odevity, monotonicity and boundedness.
We show that if $A$ satisfies a certain coherence property and the specified sparsity level is set in an appropriate range,
then AIT algorithms can find the correct support set within finite iterations.
Moreover, once the correct support set has been identified, then AIT algorithms converge to the original sparse solution exponentially fast.
In other words, the asymptotic convergence rates of AIT algorithms are linear.
It should be pointed out that the linear rates of asymptotic convergence of AIT algorithms
are not trivial since most of the thresholding operators studied in this paper are expansive.
Thus, the classical theoretical results of the Landweber iteration can not be straightly applicable to these algorithms.


The reminder of this paper is organized as follows.
In section 2, we introduce the adaptively iterative thresholding (AIT) algorithms.
In section 3, we present the main theoretical results of AIT algorithms.
In section 4, we give the proof of the main theorem.
In section 5, we discuss some related work.
We conclude the paper in section 6.

\section{Adaptively Iterative Thresholding Algorithms}

In this section,
we first give some notations used in this paper,
and then introduce the adaptively iterative thresholding algorithms.

\subsection{Notion and Notation}

For any $x\in \mathbf{R}^N$,
$x_i$ represents its $i$-th component.
Given a positive integer $k<N$, $|x_{[k]}|$ represents its $k$-th largest component of $x$ in magnitude.
For any $A\in \mathbf{R}^{M\times N}$,
$A_i \in \mathbf{R}^M$ denots its $i$th column,
$A^T$ represents its transposition.
For any index set $S$, $|S|$ denotes its cardinality, $S^c$ represents its complementary set.
Moreover, we denote by $A_S$ the submatrix of $A$ with the columns restricted to $S$.


\subsection{AIT Algorithms}

\begin{figure}
\centering
  \includegraphics[width=6.5in]{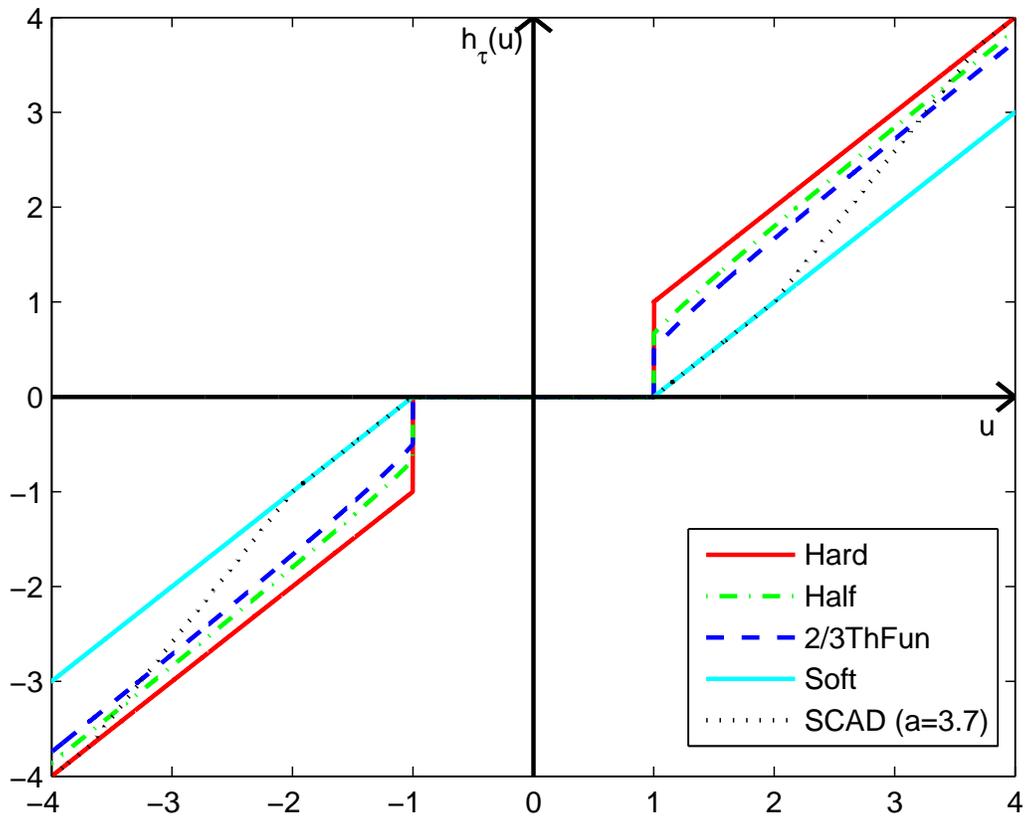}\\
  \vspace*{-20pt}
  \caption{Typical thresholding functions $h_{\tau}(u)$ with $\tau=1$.}
  \label{Fig ThFun}
  \vspace*{-30pt}
\end{figure}

The adaptively iterative thresholding algorithm for sparse solution to (\ref{Equation})
can be generally expressed as the following iterative form:
\begin{equation}
z^{(t+1)} = x^{(t)} -  A^{T}(A x^{(t)}-y),\\
\label{AIT1}
\end{equation}
\begin{equation}
x^{(t+1)} = H_{\tau^{(t+1)}}(z^{(t+1)}),
\label{AIT}
\end{equation}
where
\begin{equation}
H_{\tau^{(t+1)}}(x) = (h_{\tau^{(t+1)}}(x_1),\cdots, h_{\tau^{(t+1)}}(x_N))^T
\end{equation}
is a componentwise thresholding operator associated with a thresholding function $h_{\tau^{(t+1)}}$,
$\tau^{(t+1)}$ is the threshold value at $(t+1)$-th iteration.
More specifically, a thresholding function $h_{\tau}$ is commonly defined as
\begin{equation}
h_{\tau}(u)= \left\{
\begin{array}{cc}
f_{\tau}(u), & |u|> \tau \\
0, & {\rm otherwise}%
\end{array}%
\right.
\label{ThreshFun}
\end{equation}%
where $f_{\tau}(u)$ is formally called the defining function for any $u\in \mathbf{R}$.
We give some basic assumptions of the defining function as follows:
\begin{enumerate}
  \item \textbf{Odevity}. $f_{\tau}$ is an odd function.
  \item \textbf{Monotonicity}.  $f_{\tau}(u)\geq f_{\tau}(v)$ for any $u \geq v \geq 0$.
  \item \textbf{Boundedness}.  There exists a constant $0\leq c\leq 1$ such that $u - c \tau \leq   f_{\tau}(u) \leq u $  for $u\geq \tau$.
\end{enumerate}
The odevity and monotonicity are two regular assumptions for the defining function,
while the boundedness confines $h_{\tau}$ to be an appropriate thresholding function.
It can be noted that most of the commonly used thresholding functions satisfy these assumptions.
We list some typical examples as follows.
\\
{\bf Example 1}. Hard thresholding function for $L_{0}$ regularization ({\cite{BlumensathHard08}})
\begin{equation}
h_{\tau,0}(u)= \left\{
\begin{array}{cc}
u,  & |u|> \tau \\
0, & {\rm otherwise}%
\end{array}%
\right..
\label{HardThreshFun}
\end{equation}%
\\
{\bf Example 2}. Half thresholding function for $L_{1/2}$ regularization (\cite{L1/2TNN})
\begin{equation}
h_{\tau,1/2}(u)= \left\{
\begin{array}{cc}
{\frac{2}{3}}u
\left(1 + \cos\left({\frac{2{\pi}}{3}}-{\frac{2}{3}}\arccos\left({\frac{\sqrt{2}}{2}}{(\frac{\tau}{|u|})}^{\frac{3}{2}}\right)\right)\right),  &
|u|> \tau \\
0, & {\rm otherwise}%
\end{array}%
\right..
\label{HalfThreshFun}
\end{equation}%
\\
{\bf Example 3}. $2/3$-thresholding function for $L_{2/3}$ regularization (\cite{L2/3Cao2013})
\begin{equation}
h_{\tau,2/3}(u)= \left\{
\begin{array}{cc}
sign(u) \left( \frac{\phi_{\tau}(u)+\sqrt{\frac{2|u|}{\phi_{\tau}(u)}-\phi_{\tau}(u)^2}}{2}\right)^3,  &
|u|> \tau \\
0, & {\rm otherwise}%
\end{array}%
\right., \label{2/3ThreshFun}
\end{equation}%
where $sign(u)$ denotes as the sign function of $u$ henceforth,
$\phi_{\tau}(u) = \frac{2^{13/16}}{4\sqrt{3}}\tau^{3/16}(\cosh(\frac{\theta_{\tau}(u)}{3}))^{1/2}$
with $\theta_{\tau}(u) = arccosh(\frac{3\sqrt{3}u^2}{2^{7/4}(2\tau)^{9/8}})$.
\\
{\bf Example 4}. Soft thresholding function for $L_{1}$ regularization ({\cite{DaubechiesSoft04}})
\begin{equation}
h_{\tau,1}(u)= \left\{
\begin{array}{cc}
u - sign(u) \tau,  & |u|> \tau \\
0, & {\rm otherwise}%
\end{array}%
\right..
\label{SoftThreshFun}
\end{equation}%
\\
{\bf Example 5}. $SCAD$-thresholding function for nonconvex likelihood model ($a>2$) ({\cite{FanSCAD}})
\begin{equation}
h_{\tau,SCAD}(u)= \left\{
\begin{array}{cc}
u, &  |u|> a \tau\\
\frac{(a-1)u-sign(u)a\tau}{a-2},  &
2\tau < |u| \leq a \tau \\
u-sign(u) \tau, & \tau<|u|\leq 2\tau\\
0, & {\rm otherwise}
\end{array}%
\right..
\label{SCADThreshFun}
\end{equation}%
The plots of these thresholding functions and their corresponding boundedness parameters $c$
are shown in Figure {\ref{Fig ThFun}} and Table {\ref{TableBoundPar}}, respectively.

It can be observed that
the tuning strategies of the threshold value $\tau^{(t)}$ are crucial for AIT algorithms.
In this paper,
we consider a heuristic way for setting the threshold value,
i.e., the threshold value is set to the $(k+1)$-th largest coefficient of $z$ in magnitude at each iteration,
where $k$ is the unique algorithmic parameter and called the specified sparsity level.
Therefore, the adaptively iterative thresholding algorithms can be formulated as Algorithm 1.

It should be noticed that at $(t+1)$-th iteration,
the AIT algorithm yields a sparse solution with $k$ nonzero components by setting $\tau^{(t+1)} = |z^{(t+1)}|_{[k+1]}$ in step 4 of Algorithm 1.
To guarantee the performance of the AIT algorithm, the specified sparsity level is very critical.
Assume that the true sparsity level of the original sparse solution is $k^*$.
On one hand, when $k\geq k^*$, the results will get better as $k$ approaching to $k^*$.
On the other hand, once $k<k^*$, then the AIT algorithm fails to find the original sparse solution.
Thus, $k$ should be specified as an upper bound estimation of $k^*$.

\section{Convergence Analysis of AIT Algorithms}

In this section, we provide the convergence analysis of AIT algorithms for sparse solution to (\ref{Equation}).
For simplicity, we assume that the normalization step has been done before the analysis, that is, $\|A_{j}\|_{2}=1$ for $j=1, \ldots, N$.
We use $x^{*}=(x_{1}^{*}, \cdots, x_{N}^{*})^{T}$ to denote the original sparse solution with $k^*$ nonzeros components.
Without loss of generality, we further assume that $|x_{1}^{*}| \geq |x_{2}^{*} | \geq \cdots \geq |x_{k^*}^{*} | > 0$
and $x_{j}^{*}=0$ for $j > k^*$.
Moreover, we denote by $I^*$ and $I^{(t)}$ the support sets of $x^*$ and $x^{(t)}$, respectively.
Furthermore, we denote $I_{r}=\{1, \ldots, r\}$ for $1 \leq r \leq k^*$ as the set formed by the first $r$ largest components of $x^*$ in magnitude.
Thus, we have $I^* = I_{k^*}$.

To investigate the convergence of AIT algorithms, we introduce the coherence of a matrix $A$, which is defined as follows {\cite{DonohoCoherence}}
$$
\mu(A) = \max_{i \neq j} |\langle A_{i}, A_{j}\rangle| \quad \mbox{for} \ i, j \in \{1, \ldots, N\}.
$$
The coherence measures the maximal correlation between two different columns of $A$.
For simplicity, we use $\mu$ instead of $\mu(A)$ henceforth if there is no confusion.
In {\cite{DonohoCoherence}}, it was shown that if $k^*\leq \frac{1}{2}(1+\frac{1}{\mu})$,
then $x^*$ is the unique sparsest solution of (\ref{Equation}).
Next, we define the dynamic range of the original sparse solution as
$$Dr = \frac{\min_{i\in I^*}|x_i^*|}{\min_{i\in I^*}|x_i^*|},$$
which measures the diversity of the nonzero components of $x^*$.
Moreover, we define two positive constants in the following
\begin{equation}
T_{k^*} = k^*+ (k^*-1) \log_{(1+c)k\mu} \frac{1-(3+c)k\mu}{(3+c)-(c^2+4c+3+2/Dr)k\mu} - \log_{(1+c)k\mu}Dr,
\label{Tk*}
\end{equation}
and
\begin{equation}
T_{k^*}^* = k^*+ (k^*-1) \log_{(1+c)k^*\mu} \frac{1-(3+c)k^*\mu}{(3+c)-(c^2+4c+3+2/Dr)k^*\mu} - \log_{(1+c)k^*\mu}Dr.
\label{Tk**}
\end{equation}

With these notations, we present the main result as follows.

{\bf Theorem 1.}
Suppose that $0<\mu< \frac{1}{(3+c)k^*}$ and $k^*\leq k <\frac{1}{(3+c)\mu}$.
Then there exists a positive integer $t^*\leq T_{k^*}$ such that when $t\geq t^*$, it holds
\begin{equation}
I^* \subset I^{(t)},
\label{Th1.1}
\end{equation}
and
\begin{equation}
\|x^{(t)}-x^*\|_{\infty} \leq \frac{3+c}{2}\min_{i\in I^*}|x_i^*|\rho^{t-t^*+1}
\label{Th1}
\end{equation}
with $\rho = (1+c)k\mu<1/2.$

%
%
%
%

In Theorem 1, we justify the global convergence of AIT algorithms.
It shows that as long as $A$ satisfies a certain coherence property and the specified sparsity level $k$ is chosen in an appropriate range,
AIT algorithms can find the correct support set within finite iterations.
Furthermore, once the correct support set has been identified, then AIT algorithms converge to the original sparse solution exponentially fast.

As shown by Theorem 1 and (\ref{Tk*}),
the upper bound on the number of iterations required for identifying the correct support set is mainly determined by several parameters, i.e., $k^*$, $Dr$ and $k$.
On one hand, according to (\ref{Tk*}),
$T_{k^*}$ is monotonic increasing with respective to both $k^*$ and $Dr$.
In other words,
if the original sparse solution has more nonzero components and its dynamic range is larger,
then more iterations are commonly required to identify the correct support set.
These coincide with the common senses.
As we all known, it is generally more difficult to find a denser solution.
Also, if the dynamic range of the original solution is larger, more efforts are usually required to detect the smallest nonzero component.
On the other hand, we can easily verify that $T_{k^*}$ is monotonically increasing with respective to $k$.
Therefore, if the specified sparsity level $k$ is estimated more precisely, the number of iterations required for finding the correct support set may get fewer.
Moreover, according to (\ref{Th1}), it can be seen that AIT algorithms converge faster with smaller $\rho$ when $k$ is closer to $k^*$.
Thus, in practice, $k$ is desired to be estimated more precisely in terms of computational efficiency and convergence speed.

As analysed in the previous, a tighter upper bound estimation of the true sparsity level is more desired for the AIT algorithm in the perspectives of both theory and practice.
However, the upper bound is commonly unknown in practice.
In applications, we may conduct an empirical study or based on some known priors to yield a reasonable upper bound.
Moreover, there are several efficient ways inspired by some theoretical analysis.
In {\cite{Maleki-Donoho(2010)}}, it suggested that an upper bound can be estimated by the undersampling-sparsity
tradeoff, or ``phase-transition curve''.
However, it is generally very time-consuming to obtain the ``phase-transition curve''.
According to {\cite{Gribonval-Nielsen2003}}, it was shown that the coherence satisfies $\mu \in \left[\sqrt{\frac{N-M}{M(N-1)}},1\right]$.
The lower bound is known as the Welch bound {\cite{Welch1974}}.
Particularly, when $N\gg M$, the lower bound is approximately $\mu\geq \frac{1}{\sqrt{M}}$.
Together with Theorem 1, we can suggest $\mathcal{O}(\sqrt{M})$ as a reasonable upper bound estimation of $k^*$.

In the following, we give a corollary to show the special case with $k=k^*$.

{\bf Corollary 1.}
Suppose that $0<\mu< \frac{1}{(3+c)k^*}$ and $k=k^*$.
Then there exists a positive integer $\hat{t}^*\leq T^*_{k^*}$ such that when $t\geq \hat{t}^*$, it holds
\begin{equation}
I^* = I^{(t)},
\label{Corollary1.1}
\end{equation}
and
\begin{equation}
\|x^{(t)}-x^*\|_{\infty} \leq \frac{3+c}{2}\min_{i\in I^*}|x_i^*|\hat{\rho}^{t-\hat{t}^*+1}
\label{Corollary1}
\end{equation}
with $\hat{\rho} = (1+c)k^*\mu<1/2.$

From Corollary 1, when $k=k^*$, the AIT algorithm can recover the support set of $x^*$ exactly within finite iterations.
According to (\ref{Tk**}), it can be observed that if $k^*\mu$ is not sufficient close to $\frac{1}{3+c}$ and the dynamic range of the original sparse solutio is not too large,
then the log term about $k^*\mu$ and $Dr$ in the second and third terms of (\ref{Tk**}) respectively are relatively small constants.
In this case, the number of iterations required for the AIT algorithm is about several times of $k^*$.
For an instance,
assume that $k^*=9$, $\mu=\frac{1}{40}$ and $Dr=10$,
according to (\ref{Tk**}),
the number of iterations required for $hard$, $soft$ and $half$ algorithms are 20, 42 and 25,
which are about 2, 5 and 3 times of $k^*$, respectively.
Motivated by this observation, we can suggest an efficient halting rule for AIT algorithms through setting the number of maximum iterations according to the true sparsity level.

It can be observed from Corollary 1 that the boundedness parameter $c$ plays an important role in the guarantees of the convergence of AIT algorithms.
The restriction of the matrix $A$ gets stricter as $c$ increasing.
As shown in Table 1, among these AIT algorithms,
hard algorithm permits the weakest requirement of $A$ with $\mu<\frac{1}{3k^*}$,
while soft algorithm requires the strictest restriction of $A$ with $\mu<\frac{1}{4k^*}$.
It should be noticed that the restriction on $\mu$ is relatively loose and can be attained in practice.
In fact, it was shown that the coherence $\mu$ is in the order of $\sqrt{\log N /M}$
for the random matrix where entries of $A$ are independently and identically gaussian distributed \cite{Candes-and-Plan(2009)}.
This implies that $k^* = O(M^{\xi_{1}})$ might suffice for the AIT algorithm
when $\log N = O (M^{\xi_{2}})$ for some positive constants $\xi^{1}$ and $\xi^{2}$ satisfying $2\xi^{1} + \xi^{2} < 1$.

{\bf Remark 1.} As shown by the proof of Theorem 1 in Section 4, it is interested that
the procedure of identifying the correct support set is a sequential recruitment process.
That is, the supports are sequentially recruited in a descending order of the values of their coefficients
with the larger one being identified not later than the smaller one.
This procedure may be very useful to certain applications such as feature screening problem in statistics.

\section{ Proof of Theorem 1}

We denote $i_{[k+1]}^{(t)} = \arg \min_{i\in \{1,2,\cdots,N\}} \left\{i: \left|z_{i}^{(t)}\right| = \left|z^{(t)}\right|_{[k+1]}\right\}$
and then let $\Lambda_{[k+1]}^{(t)} = I^{(t)}\cup \left\{i_{[k+1]}^{(t)}\right\}$.
To prove Theorem 1, we need the following lemmas.
First, we give a lemma to bound the gap between the components of $x^{(t)}$ and $z^{(t)}$ at $t$-th iteration,
which is served as the basis of the other lemmas.

{\bf Lemma 1.} At any $t$-th iteration ($t\geq 1$), there exists an $i_0^{(t)} \in \Lambda_{[k+1]}^{(t)}\setminus I^*$, such that

(i) for any $i\in I^{(t)}$,
\begin{equation}
\left|z_i^{(t)}-x_i^{(t)}\right| \leq c \left|z_{i_0^{(t)}}^{(t)} - x_{i_0^{(t)}}^*\right|,
\label{Lemma1.1}
\end{equation}
where $c$ is the boundedness parameter of the associated thresholding function;

(ii) for any $i\notin I^{(t)}$,
\begin{equation}
\left|z_i^{(t)}-x_i^{(t)}\right| \leq  \left|z_{i_0^{(t)}}^{(t)} - x_{i_0^{(t)}}^*\right|.
\label{Lemma1.2}
\end{equation}
Here, it should be mentioned that $ x_{i_0^{(t)}}^*=0$ and we keep it in (\ref{Lemma1.1}) and (\ref{Lemma1.2}) only for better formats.

{\bf Proof.}
(i) For $i\in I^{(t)}$, by the definition of the thresholding function $H_{\tau}$ and the boundness assumption of $f_{\tau}$,
it holds
\begin{equation}
\left|z_i^{(t)} - x_{i}^{(t)}\right| = \left| z_{i}^{(t)}-f_{\tau^{(t)}}(z_{i}^{(t)})\right| \leq c \tau^{(t)} = c \left|z^{(t)}\right|_{[k+1]}.
\label{Lemma1.3}
\end{equation}
Since $i_{[k+1]}^{(t)} \notin I^{(t)}$, then the cardinality of $\Lambda_{[k+1]}^{(t)}$ is $k+1$.
Moreover, by $|I^*|=k^*<k+1$,
then there exists an index $i_0^{(t)}$ such that $i_0^{(t)}\in \Lambda_{[k+1]}^{(t)} \setminus I^*$.
Thus, (\ref{Lemma1.3}) becomes
\begin{equation}
\left|z_i^{(t)} - x_i^{(t)}\right| \leq c\left|z^{(t)}\right|_{[k+1]} \leq c\left|z_{i_0^{(t)}}^{(t)}\right| = c\left|z_{i_0^{(t)}}^{(t)}-x_{i_0^{(t)}}^{*}\right|.
\label{Lemma2.4}
\end{equation}

(ii) Similarly, for any $i\notin I^{(t)}$, it holds
\begin{equation}
\left|z_i^{(t)} - x_i^{(t)}\right| = \left|z_i^{(t)}\right| \leq \left|z^{(t)}\right|_{[k+1]} \leq \left|z_{i_0^{(t)}}^{(t)}-x_{i_0^{(t)}}^{*}\right|.
\label{Lemma2.5}
\end{equation}
Thus, we end the proof of this lemma.

In the next, we give a lemma to show that the largest component (in magnitude) of $x^*$ will be detected at the first iteration.

{\bf Lemma 2.} Suppose that $0<\mu <\frac{1}{2k^*-1}$ and $k^*\leq k<\frac{1}{2}(1+\frac{1}{\mu})$.
Then at the first iteration, it holds:

(i) $\{1\}\subset I^{(1)}$;

(ii) for any $j\in I^{(1)}$,
$
\left|x_j^{(1)} - x_j^*\right|\leq \frac{(1+c)(3+c)}{2} k\mu \left|x_1^*\right|.
$

{\bf Proof.}
First, we show that $\{1\} \subset I^{(1)}$.
On one hand, we observe that
\begin{equation*}
\left|z_1^{(1)}\right|
= \left|x_1^* + \sum_{i\in I^*\setminus {\{1\}}} \langle A_1, A_i\rangle x_i^*\right|
\geq |x_1^*| - \mu \sum_{i=2}^{k^*} |x_i^*|
\geq |x_1^*| - (k-1) \mu |x_1^*|.
\end{equation*}
On the other hand, for any $i\notin I^*$, it holds
\begin{equation*}
\left|z_i^{(1)}\right|
= \left|\sum_{j=1}^{k^*} \langle A_i, A_j \rangle x_j^* \right|
\leq {k^*}\mu \left|x_1^*\right|
\leq k \mu \left|x_1^*\right|.
\end{equation*}
Since $k<\frac{1}{2}(1+\frac{1}{\mu})$, then $k \mu \left|x_1^*\right| < \left|x_1^*\right| - (k-1) \mu \left|x_1^*\right|,$
which implies that
\begin{equation*}
\left|z_1^{(1)}\right| > \max_{i\notin I^*} \left|z_i^{(1)}\right|.
\end{equation*}
Thus, $\{1\} \subset I^{(1)}$.

Next, we give the error bound.
For any $j\in I^{(1)}$, we observe that
\begin{equation}
\left|x_j^{(1)} - x_j^*\right|
\leq \left|x_j^{(1)} - z_j^{(1)}\right| + \left|z_j^{(1)} - x_j^*\right|
\leq c\left|x_{i_0^{(1)}}^{*} - z_{i_0^{(1)}}^{(1)}\right| + \left|z_j^{(1)} - x_j^*\right|,
\label{Lemma2.5}
\end{equation}
where the second inequality holds for Lemma 1.
Furthermore, for any $i$, it holds
\begin{equation}
\left|z_i^{(1)} - x_i^*\right| = \left|\sum_{j\in I^*\setminus {\{i\}}} \langle A_i, A_j \rangle x_j^* \right| \leq k^* \mu \left|x_1^*\right| \leq k\mu \left|x_1^*\right|.
\label{Lemma2.6}
\end{equation}
Combining (\ref{Lemma2.5}) with (\ref{Lemma2.6}), for any $j\in I^{(1)}$, it holds
\begin{equation*}
\left|x_j^{(1)} - x_j^*\right|
\leq (1+c) k\mu \left|x_i^*\right|
\leq \frac{(1+c)(3+c)}{2} k\mu \left|x_1^*\right|.
\end{equation*}
Thus, we end the proof of this lemma.

{\bf Lemma 3.}
Suppose that $0<\mu<\frac{1}{(3+c)k^*}$ and $k^*\leq k< \frac{1}{(3+c)\mu}$.
Moreover, assume that at $m$-th iteration, $I_r\subset I^{(m)}$ ($0<r\leq k^*$)
and for any $j\in I^{(m)}$, it holds $\left|x_j^{(m)} - x_j^*\right| \leq \frac{(1+c)(3+c)}{2} k\mu \left|x_r^*\right|$.
Then at $(m+s)$-th iteration ($s\geq 1$), it holds

(i) for any $j$,
\begin{equation}
\left|z_j^{(m+s)} -x_j^*\right|
\leq \frac{(3+c)}{2} k\mu\left((1+c)k\mu\right)^s \left|x_r^*\right| + k\mu \left|x_{r+1}^*\right|\left[1+(1+c)k\mu+\cdots+ ((1+c)k\mu)^{s-1}\right];
\nonumber
\end{equation}

(ii) for any $i\in I^{(m+s)}$,
\begin{equation}
\left|x_i^{(m+s)} -x_i^*\right| \leq
\frac{(3+c)}{2} ((1+c)k\mu)^{s+1} \left|x_r^*\right| + k\mu \left|x_{r+1}^*\right|\left[(1+c)k\mu+\cdots+ ((1+c)k\mu)^{s}\right] ;
\nonumber
\end{equation}

(iii) $I_r \subset I^{(m+s)}$.

{\bf Proof.}
We prove this lemma by induction.
First, when $s=1$, for any $i\in I^{(m+1)}$, it holds
\begin{equation*}
\left|x_i^{(m+1)} - x_i^*\right| \leq \left|x_i^{(m+1)} -z_i^{(m+1)}\right| + \left|z_i^{(m+1)} - x_i^*\right|.
\end{equation*}
By Lemma 1, there exists an $i_0^{(m+1)}\in \Lambda_{[k+1]}^{(m+1)}\setminus I^*$ such that
$$
\left|x_i^{(m+1)} -z_i^{(m+1)}\right| \leq c\left|z_{i_0^{(m+1)}}^{(m+1)} - x_{i_0^{(m+1)}}^*\right|,
$$
then it holds
\begin{equation}
\left|x_i^{(m+1)} - x_i^*\right| \leq c\left|z_{i_0^{(m+1)}}^{(m+1)} - x_{i_0^{(m+1)}}^*\right| + \left|z_i^{(m+1)} - x_i^*\right|.
\label{Lemma3.1}
\end{equation}
Moreover, for any  $j$, it holds
\begin{eqnarray}
\left|z_j^{(m+1)} - x_j^*\right|
&=& \left|\sum_{i\in I^{(m)}\cup I^*\setminus {\{j\}}} \langle A_j,A_i \rangle (x_i^*- x_i^{(m)})\right| \nonumber\\
&=& \left|\sum_{i\in I^{(m)}\setminus {\{j\}}} \langle A_j,A_i \rangle (x_i^*- x_i^{(m)})
+ \sum_{i\in I^*\setminus (I^{(m)}\cup {\{j\}})} \langle A_j,A_i \rangle x_i^* \right| \nonumber\\
&\leq& k\mu\left(\frac{(1+c)(3+c)}{2} k\mu \left|x_r^*\right|\right) + (k^*-r) \mu \left|x_{r+1}^*\right| \nonumber\\
&\leq& \frac{(3+c)}{2}k\mu \left((1+c)k\mu \left|x_r^*\right|\right) + k\mu \left|x_{r+1}^*\right|.
\label{Lemma3.2}
\end{eqnarray}
Combining (\ref{Lemma3.1}) with (\ref{Lemma3.2}), for any $i\in I^{(m+1)}$, it holds
\begin{eqnarray}
\left|x_i^{(m+1)} - x_i^*\right|
&\leq& (1+c) \left(\frac{(3+c)}{2}k\mu ((1+c)k\mu \left|x_r^*\right|) + k\mu \left|x_{r+1}^*\right|\right) \nonumber\\
&=& \frac{(3+c)}{2} ((1+c) k\mu)^2 \left|x_r^*\right| + (1+c) k\mu \left|x_{r+1}^*\right|.
\label{Lemma3.3}
\end{eqnarray}
Then we need to prove that $I_r \subset I^{(m+1)}$.
According to (\ref{Lemma3.2}), for any $j$, it holds
\begin{equation*}
|z_j^{(m+1)} - x_j^*| \leq \left(1+\frac{(3+c)(1+c)}{2}k\mu \right) k\mu |x_r^*|.
\end{equation*}
Since $k<\frac{1}{(3+c)\mu}$, it holds
\begin{equation*}
\left(1+\frac{(3+c)(1+c)}{2} k\mu\right) k\mu  < \frac{1}{2}.
\end{equation*}
Then for any $j$, it holds
\begin{equation}
\left|z_j^{(m+1)} - x_j^*\right| < \frac{1}{2}\left|x_r^*\right|.
\label{Lemma3.6}
\end{equation}
According to (\ref{Lemma3.6}), we observe that, for any $i\in I_r$,
\begin{equation}
\left|z_i^{(m+1)}\right| \geq \left|x_i^*\right| - \left|z_i^{(m+1)} - x_i^*\right| \geq \left|x_r^*\right| - \frac{1}{2}\left|x_r^*\right| > \frac{1}{2}\left|x_r^*\right|.
\label{Lemma3.7}
\end{equation}
While for $i\notin I^*$,
\begin{equation}
\left|z_i^{(m+1)}\right| = \left|z_i^{(m+1)} -x_i^*\right| < \frac{1}{2}\left|x_r^*\right|.
\label{Lemma3.8}
\end{equation}
With (\ref{Lemma3.7}) and (\ref{Lemma3.8}), it follows that $I_r \subset I^{(m+1)}$.
Therefore, the conclusion holds for $s=1$.

Second, assume that the conclusion holds for $s$ ($s\geq 1$), then we need to check it holds for $s+1$.
The proof is similar to the case $s=1$ and we omit it here.

{\bf Lemma 4.} Suppose that $0<\mu<\frac{1}{(3+c)k^*}$ and $k^*\leq k< \frac{1}{(3+c)\mu}$.
Moreover, assume that at $m$-th iteration, $I_r\subset I^{(m)}$ ($r<k^*$) and for any $j\in I^{(m)}$,
$|x_j^{(m)}-x_j^*| \leq \frac{(1+c)(3+c)}{2} k\mu |x_r^*|$.
Then it holds:

(i) the index $\{r+1\}$ will be detected after at most $l_r$ iterations with
$$
l_r = \left\lfloor \log_{(1+c)k\mu} \frac{1-(3+c)k\mu}{(3+c)(1-(1+c)k\mu)|x_r^*|/|x_{r+1}^*| - 2k\mu}\right\rfloor,
$$
where the function $\lfloor u \rfloor$ denotes the smallest integer not less than $u$ for any $u\in \mathbb{R}$.

(ii) for any $j\in I^{(m+l_r+1)}$,
$$
\left|x_j^{(m+l_r+1)} - x_j^*\right| < \frac{(1+c)(3+c)}{2} k\mu \left|x_{r+1}^*\right| .
$$

{\bf Proof.}
We first show that the index $\{r+1\}$ will be detected after at most $l_r$ iterations, and then give the error bound.
According to Lemma 3, at $(m+l_r)$-th iteration, for any $j$, it holds
\begin{eqnarray}
\left|z_j^{(m+l_r)} - x_j^*\right|
&\leq& \frac{(3+c)}{2} ((1+c)k\mu)^{l_r} \left|x_r^*\right| + k\mu \left|x_{r+1}^*\right|\left(1+\cdots + ((1+c)k\mu)^{l_r-1}\right)
 \nonumber\\
&<& \frac{(3+c)}{2} ((1+c)k\mu)^{l_r} \left|x_r^*\right| + k\mu \left|x_{r+1}^*\right| \frac{1-((1+c)k\mu)^{l_r}}{1-(1+c)k\mu}
\nonumber\\
&=& \left|x_{r+1}^*\right| \left(\frac{(3+c)}{2} ((1+c)k\mu)^{l_r} \frac{\left|x_r^*\right|}{\left|x_{r+1}^*\right|} + k\mu \frac{1-((1+c)k\mu)^{l_r}}{1-(1+c)k\mu}\right)
\nonumber\\
&\leq& \left|x_{r+1}^*\right|\left(\frac{(3+c)}{2} ((1+c)k\mu)^{l_r} \frac{\left|x_r^*\right|}{\left|x_{r+1}^*\right|} + k\mu \frac{1-((1+c)k\mu)^{l_r}}{1-(1+c)k\mu}\right).
\nonumber
\end{eqnarray}
Since
$$
l_r
\geq \log_{(1+c)k\mu} \frac{1-(3+c)k\mu}{(3+c)(1-(1+c)k\mu)|x_r^*|/|x_{r+1}^*| - 2k\mu},
$$
then
\begin{equation*}
\frac{(3+c)}{2} ((1+c)k\mu)^{l_r} \frac{|x_r^*|}{|x_{r+1}^*|} + k\mu \frac{1-((1+ck\mu)^{l_r}}{1-(1+ck\mu)}
\leq \frac{1}{2}.
\end{equation*}
Thus, for any $j$, it holds
\begin{equation}
\left|z_j^{(m+l_r)} - x_j^*\right| <\frac{1}{2}\left|x_{r+1}^*\right|.
\label{Lemma4.3}
\end{equation}
By (\ref{Lemma4.3}), on one hand
\begin{equation}
\left|z_{r+1}^{(m+l_r)}\right| \geq \left|x_{r+1}^*\right| - \left|z_{r+1}^{(m+l_r)} - x_{r+1}^*\right| > \frac{1}{2} \left|x_{r+1}^*\right|,
\label{Lemma4.4}
\end{equation}
and on the other hand, for any $j\notin I^*$,
\begin{equation}
|z_j^{(m+l_r)}| = |z_j^{(m+l_r)} - x_j^*| < \frac{1}{2}|x_{r+1}^*|.
\label{Lemma4.5}
\end{equation}
With (\ref{Lemma4.4}) and (\ref{Lemma4.5}), it shows that
$\{r+1\}$ will be detected at $(m+l_r)$-th iteration, that is, $\{r+1\} \subset I^{(m+l_r)}$.

Next, we give the upper bound of the error.
For any $i\in I^{(m+l_r+1)}$, it holds
\begin{eqnarray}
\left|x_i^{(m+l_r+1)} - x_i^*\right|
&=& \left|\sum_{j\in I^{(m+l_r)}\setminus {\{i\}}} \langle A_i,A_j \rangle (x_j^* - x_j^{(m+l_r)})
+ \sum_{j\in I^*\setminus (I^{(m+l_r)}\cup {\{i\}})} \langle A_i,A_j \rangle \beta_j^* \right| \nonumber\\
&\leq& \mu \sum_{j\in I^{(m+l_r)}\setminus {\{i\}}} \left|x_j^* - x_j^{(m+l_r)}\right| + (k^*-r-1)\mu \left|x_{r+1}^*\right| .
\label{Lemma4.6}
\end{eqnarray}
Moreover, for any $j\in I^{(m+l_r)}$, it holds
\begin{equation}
\left|x_j^*-x_j^{(m+l_r)}\right| \leq \left|x_j^*-z_j^{(m+l_r)}\right| + \left|z_j^{(m+l_r)} - x_j^{(m+l_r)}\right|.
\label{Lemma4.7}
\end{equation}
According to Lemma 1  and (\ref{Lemma4.3}), then (\ref{Lemma4.7}) becomes
\begin{equation}
\left|x_j^*-x_j^{(m+l_r)}\right|
< \frac{1}{2}\left|x_{r+1}^*\right| + c\left|z_{i_0^{(m+l_r)}}^{(m+l_r)} - x_{i_0^{(m+l_r)}}^*\right|
< \frac{1+c}{2} \left|x_{r+1}^*\right|.
\label{Lemma4.8}
\end{equation}
Combining (\ref{Lemma4.6}) and (\ref{Lemma4.8}), for any $i\in I^{(m+l_r+1)}$, it holds
\begin{eqnarray}
\left|x_i^{(m+l_r+1)} - x_i^*\right|
&\leq& \frac{(1+c)}{2}k\mu \left|x_{r+1}^*\right| + (k^*-r-1)\mu \left|x_{r+1}^*\right|  \nonumber\\
&=& \left(\frac{1+c}{2} + \frac{k^*-r-1}{k}\right) k\mu \left|x_{r+1}^*\right|\nonumber\\
&\leq& \frac{(1+c)(3+c)}{2} k\mu \left|x_{r+1}^*\right|.
\nonumber
\end{eqnarray}
Therefore, for any $i\in I^{(m+l_r+1)}$, it holds
\begin{equation*}
\left|x_i^{(m+l_r+1)} - x_i^*\right| \leq \frac{(1+c)(3+c)}{2} k\mu \left|x_{r+1}^*\right|.
\end{equation*}
Thus, we end the proof of Lemma 4.

{\bf Proof of Theorem 1.}
With these lemmas, we prove Theorem 1 inductively.
For $i=1$, by Lemma 2, the largest component (in magnitude) will be detected at the first iteration, that is, $I_1 = \{1\} \subset I^{(1)}$. By Lemma 3, once the first largest index is identified, then it remains in the support set forever.
Furthermore, by Lemma 4, the second largest component will be identified after at most $l_1$ iterations, i.e., $I_2 \subset I^{(t)}$ when $t\geq 1+l_1$. In order to obtain the required error bound for the inductive procedure, one more iteration should be implemented.
When this procedure is repeated for $r$ times with $0<r\leq k^*-1$, it holds $I_{r+1}\subset I^{(t)}$ when $t\geq r + \sum_{i=1}^{r-1} l_i$.
Furthermore, by Lemma 3, once all the correct indices are detected, they remains in the support set and the error estimation of the iteration can be obtained.
Therefore, there exists an integer constant $t^*\leq k^* + \sum_{i=1}^{k^*-1} l_i$ such that when $t\geq t^*$,
it holds $I^*\subset I^{(t)}$ and the error estimation of the iteration can be achieved.
Moreover, by the definition of $l_i$ in Lemma 4 and the fact that $|x^*_i|/|x^*_{i+1}|\leq Dr$, it holds
\begin{eqnarray}
l_i
&\leq& \log_{(1+c)k\mu} \frac{1-(3+c)k\mu}{(3+c)(1-(1+c)k\mu)|x_i^*|/|x_{i+1}^*| - 2k\mu} \nonumber\\
&\leq& \log_{(1+c)k\mu} \frac{1-(3+c)k\mu}{(3+c)-(c^2+4c+3+2/Dr)k\mu} -\log_{(1+c)k\mu} \frac{|x_i^*|}{|x_{i+1}^*|}
\label{li}
\end{eqnarray}
for $i=1,\cdots,k^*-1$. Therefore,
\begin{equation*}
k^*+\sum_{i=1}^{k^*-1} l_i \leq k^*+(k^*-1)\log_{(1+c)k\mu} \frac{1-(3+c)k\mu}{(3+c)-(c^2+4c+3+2/Dr)k\mu} - \log_{(1+c)k\mu} \frac{|x_1^*|}{|x_{k^*}^*|}=T_{k^*}.
\end{equation*}

Thus, we obtain the proof of Theorem 1.

\section{Related Work}

In this section, we first discuss some related work of AIT algorithms,
and then give some comparisons with other typical algorithms including BP, OMP, CoSaMP in terms of the sufficient condition for convergence and computational complexity.

{\bf (i) On related work of AIT algorithms.}
In {\cite{MalekiITA2009}}, Maleki provided some similar results for two special AIT algorithms, i.e., the hard and soft algorithms
with $k=k^*$.
The sufficient conditions for convergence are $\mu<\frac{1}{3.1k^*}$ and  $\mu<\frac{1}{4.1k^*}$ for hard and soft algorithms, respectively.
As shown by Corollary 1, our conditions for both algorithms are slightly weaker than Maleki's conditions.
Moreover, from Theorem 1, we show the robustness of AIT algorithms to the specified sparsity levels,
which is very important in practice.
Except the hard and soft algorithms, as far as we know,
there are no similar results on the global convergence of other AIT algorithms such as half, SCAD and MCP algorithms for sparse solution to the underdetermined linear equations.

Besides the coherence property, another important property called the restricted isometry property (RIP) is commonly used to characterize the performance of an algorithm for sparse solution to (\ref{Equation}).
The $s$-order restricted isometry constant (RIC), $\delta_s$ of $A$ is defined as the smallest constant $0<\delta<1$ such that
\begin{equation}
(1-\delta) \|x\|_2^2 \leq \|Ax\|_2^2 \leq (1+\delta)\|x\|_2^2, ~\forall \|x\|_0\leq s.
\label{RIP}
\end{equation}
In {\cite{CaiCoherence}}, it was demonstrated that if $A$ has unit-norm columns and coherence $\mu$, then $A$ has the $(s,\delta_s)$-RIP with
\begin{equation}
\delta_s \leq (s-1)\mu.
\label{RIP-Coh}
\end{equation}
In terms of RIP, Blumensath and Davies justified the performance of the hard algorithm when applied to signal recovery problem {\cite{Blumensath08CS}}.
It was shown that if $A$ satisfies a certain RIP with $\delta_{3k^*}<\frac{1}{8\sqrt{2}-1}$,
then the global convergence of the hard algorithm can be guaranteed.
Later, this condition was significantly improved to by Foucart {\cite{FoucartRIP2010}}, i.e., $\delta_{3k^*}<\frac{1}{2}$.
Together with (\ref{RIP-Coh}), we can easily deduce a coherence based sufficient condition of convergence, that is, $\mu<\frac{1}{2(3k^*-1)}$.
As compared with the existing RIP based conditions,
it is hard to claim whether our conditions are better.
Instead, we can give some useful remarks on these conditions.
On one hand, the sufficient conditions based on coherence can be in general verified much easier than those based on RIP.
On the other hand, the RIP based conditions can be generalized and improved usually easier than those based on coherence.

{\bf (ii) On comparison with other algorithms.}
For better comparison, we list the state-of-the-art results on sufficient conditions of some typical algorithms including BP, OMP, CoSaMP, hard, soft, half and other AIT algorithms in Table {\ref{TableERC}}.

From Table {\ref{TableERC}}, in the perspective of coherence, the sufficient conditions of AIT algorithms are slightly stricter than those of BP and OMP algorithms.
However, AIT algorithms are generally faster than both algorithms with lower computational complexities, especially for large scale applications.
As analyzed in Section 3, the number of iterations required for the convergence of the AIT algorithm is empirically of the same order of the original sparsity level $k^*$, that is, $\mathcal{O}(k^*)$.
At each iteration of the AIT algorithm, only some simple matrix-vector multiplications and a projection on the vector need to be done,
and thus the computational complexity per iteration is $\mathcal{O}(MN)$.
Therefore, the total computational complexity of the AIT algorithm is $\mathcal{O}(k^*MN)$.
While the total computational complexities of BP and OMP algorithms are generally $\mathcal{O}(M^2N)$ and $\max\{\mathcal{O}(k^*MN), \mathcal{O}(\frac{(k^*)^2(k^*+1)^2}{4})\}$, respectively.
It should be pointed out that the computational complexity of OMP algorithm is related to the commonly used halting rule of OMP algorithm,
that is, the number of maximal iterations is set to be the true sparsity level $k^*$.

As another important greedy algorithm, CoSaMP algorithm identifies multicomponents (commonly $2k^*$) at each iteration.
From Table \ref{TableERC}, the RIP based sufficient condition of CoSaMP is $\delta_{4k^*}<0.384$ and a deduced coherence based sufficient condition is $\mu<\frac{0.384}{4k^*-1}$.
In the perspective of coherence, our conditions for AIT algorithms are better than CoSaMP, though this comparison is not very reasonable.
At each iteration of CoSaMP algorithm, some simple matrix-vector multiplications and a least squares problem should be considered.
Thus, the computational complexity per iteration of CoSaMP algorithm is generally $\max\{\mathcal{O}(MN), \mathcal{O}((3k^*)^3)\}$,
which is higher than those of AIT algorithms, especially when $k^*$ is very large.
However, the number of iterations required for CoSaMP algorithm is commonly fewer than those of AIT algorithms,
since the speed of convergence of CoSaMP algorithm is exponential while those of AIT algorithms are asymptotically exponential,
that is, AIT algorithms converge exponentially fast after certain iterations.
Therefore, as claimed in the introduction, when applied to very sparse case, both OMP and CoSaMP algorithms may be more efficient than AIT algorithms.
While AIT algorithms may be better when applied to more general cases.

\section{Conclusion}

In this paper, we provide the convergence analysis of a wide class of adaptively iterative thresholding (AIT) algorithms
for sparse solution to an underdetermined system of linear equations $y=Ax$.
We prove that as long as $A$ satisfies a certain coherence property and the specified sparsity level is set in an appropriate range,
AIT algorithms can identify the correct support set within finite steps.
Furthermore, we demonstrate that the asymptotic convergence rates of AIT algorithms are linear,
that is, once the correct support set has been identified, AIT algorithms converge to the original sparse solution exponentially fast.
It is interested that the procedure of finding the correct support set is a sequential recruitment process,
i.e., the supports are sequentially recruited into the support set in the descending order of the magnitudes of their coefficients.
This property may be very useful to certain applications such as feature screening problem.
It should be noted that most of the commonly used iterative thresholding algorithms (say, hard, soft, half and SCAD algorithms) are included in the class of iterative thresholding algorithms studied in this paper.
Besides the hard and soft algorithms, we provide some fundamental guarantees on the performance of the other AIT algorithms for sparse solution to an underdetermined linear equations.


\bibliographystyle{elsarticle-num}

\begin{thebibliography}{80}

\bibitem{Donoho06}
D. L. Donoho, Compressed sensing,
IEEE Transactions on Information Theory, 52 (4): 1289-1306, 2006.

\bibitem{Candes06}
 E. J. Candes, J. Romberg, and T. Tao,
Robust uncertainty principles: exact signal reconstruction from highly incomplete frequency information,
 IEEE Transactions on Information Theory, 52 (2): 489-509, 2006.

\bibitem{MallatMP1993}
S. Mallat and Z. Zhang,
Matching pursuits with time-frequency dictionaries,
IEEE Transactions on Signal Processing, 41 (12): 3397-3415, 1993.

\bibitem{PatiOMP1993}
Y. Pati, R. Rezaifar and P. Krishnaprasad,
Orthogonal matching pursuit: recursive function approximatin with applications to wavelet decomposition,
In Asilomar Conf. Signals, Syst., Comput., Pacific Grove, CA, 1993.


\bibitem{TroppOMP2007}
J. A. Tropp and A. Gilbert,
Signal recovery from random measurements via orthogonal mathching pursuit,
IEEE Transactions on Information Theory, 2007, 53: 4655-4666.


\bibitem{DonohoStOMP}
D. L. Donoho, Y. Tsaig, O. Drori and J.-L. Starck,
Sparse solution of underdetermined systems of linear equations by stagewise orthogonal matching pursuit,
IEEE Transactions on Information Theory, 58 (2): 1094 - 1121, 2012.


\bibitem{Needell2010ROMP}
D. Needell and R. Vershynin, Signal recovery from incomplete and inaccurate measurements via Regularized Orthogonal Matching
Pursuit, IEEE Journal of Selected Topics in Signal Processing, 4: 310-316, 2010.


\bibitem{CoSaMP}
D. Needell and J. A. Tropp,
CoSaMP: Iterative signal recovery from incomplete and inaccurate samples,
Applied and Computational Harmonic Analysis, 26 (3): 301-321, 2008.


\bibitem{DaiSP}
W. Dai and O. Milenkovic,
Subspace pursuit for compressive sensing signal recontruction,
IEEE Transactions on Information Theory, 55 (5): 2230-2249, 2009.


\bibitem{TroppGreedyAlg}
J. A. Tropp and S. Wright,
Computational methods for sparse solution of linear inverse problems,
in: Proceedings of the IEEE, 98: 948-958, 2010.

\bibitem{Donoho98}
S. S. Chen, D. L. Donoho, and M. A. Saunders,
Atomic decomposition by basis pursuit, SIAM Journal on Scientific Computing, 20: 33-61, 1998.


\bibitem{Chartrand2007}
R. Chartrand, Exact reconstruction of sparse signals via nonconvex minimization.
IEEE Signal Processing Letters, 14 (10): 707-710, 2007.


\bibitem{Chartrand2008}
R. Chartrand and V. Staneva,
Restricted isometry properties and nonconvex compressive sensing, Inverse Problems, 24: 1-14, 2008.


\bibitem{XuHalf2012}
Z. B. Xu, H. Zhang, Y. Wang, X. Y. Chang and Y. Liang,
$L_{1/2}$ regularizater. Science in China, series F-Information Science, 53: 1159-1169, 2010.


\bibitem{L1/2TNN}
Z. B. Xu, X.Y. Chang, F. M. Xu and H. Zhang,
 $L_{1/2}$ regularization: a thresholding representation theory and a fast solver,
IEEE Transactions on Neural Networks and Learning Systems, 23: 1013-1027, 2012.


\bibitem{Candes2008RL1}
E. J. Candes, M. B. Wakin and S. P. Boyd,
Enhancing sparsity by reweighted $l_1$ minimization,
Journal of Fourier Analysis and Applications, 14 (5): 877-905, 2008.

\bibitem{FanSCAD}
J. Fan, J. and R. Li,
Variable selection via nonconcave penalized likelihood and its oracle properties, Journal of the American
Statistical Association, 96: 1348-1360, 2001.

\bibitem{ZhangMCP2010}
C. H. Zhang, Nearly unbiased variable selection under minimax concave penalty,
The Annals of Statistics, 38 (2): 894-942, 2010.


\bibitem{FOCUSS}
I. F. Gorodnitsky and B. D. Rao,
Sparse signal reconstruction from limited data using FOCUSS: a re-weighted minimum norm algorithm,
IEEE Transactions on Signal Processing, 45 (3): 600-616, 1997.

\bibitem{DaubechiesIRLS}
I. Daubechies, R. Devore, M. Fornasier and C. S. Gunturk,
Iteratively reweighted least squares minimization for sparse recovery,
Communications on Pure and Applied Mathematics, 63: 1-38, 2010.


\bibitem{DaubechiesSoft04}
I. Daubechies, M. Defries and C. De Mol,
An iterative thresholding algorithm for linear inverse problems with a sparisity constraint,
Communications on Pure and Applied Mathematics, 57: 1413-1457, 2004.


\bibitem{L2/3Cao2013}
W. F. Cao, J. Sun and Z. B. Xu, Fast image deconvolution using closed-form thresholding formulas of $L_q$ ($q=1/2,2/3$) regularization,
Journal of Visual Communication and Image Representation, 24: 31-41, 2013.


\bibitem{BlumensathHard08}
T. Blumensath and M. E. Davies,
Iterative thresholding for sparse approximation, Journal of Fourier Analysis and Application, 14 (5): 629-654, 2008.


\bibitem{Qian20011}
Y. T. Qian, S. Jia, J. Zhou and A. Robles-Kelly,
Hyperspectral unmixing via $L_{1/2}$ sparsity-constrained nonnegative matrix factorization,
IEEE Transactions on Geoscience and Remote Sensing, 49 (11): 4282-4297, 2011.

\bibitem{Zeng20012SAR}
J. S. Zeng, J. Fang, Z. B. Xu,
Sparse SAR imaging based on $L_{1/2}$ regularization,
Science  China Information Sciences, 55: 1755-1775, 2012.

\bibitem{Zeng20013AccSAR}
J. S. Zeng, Z. B. Xu, B. C. Zhang, W. Hong, Y. R. Wu.
Accelerated $L_{1/2}$ regularization based SAR imaging via BCR and reduced Newton skills,
Signal Processing, 93: 1831-1844, 2013.


\bibitem{Blumensath08CS}
T. Blumensath and M. E. Davies,
Iterative hard thresholding for compressed sensing,
Applied and Computational Harmonic Analysis, 27: 265-274, 2008.


\bibitem{MalekiITA2009}
A. Maleki, Coherence analysis of iteative thresholding algorithms,
in Forty-Seventh Annual Allerton Conference, Allerton House, UIUC, Illinois, USA, 2009.


\bibitem{DonohoCoherence}
D. L. Donoho and M. Elad,
Optimally sparse representation in general (nonorthogonal) dictionaries via $l_1$ minimization,
Proceedings of the National Academy of Sciences, 100 (5): 2197-2202, 2003.


\bibitem{Maleki-Donoho(2010)}
A. Maleki and D. L. Donoho,
Optimally tuned iterative reconstruction algorithms for compressed sensing,
IEEE Journal of Selected Topics in Signal Processing, 4(2): 330-341, 2010.

\bibitem{Gribonval-Nielsen2003}
R. Gribonval and M. Nielsen,
Sparse representations in unions of bases,
IEEE Transactions on Information Theory,
49 (12): 3320-3325, 2003.


\bibitem{Welch1974}
L. R. Welch,
Lower bounds on the maximum cross correlation of signals,
IEEE Transaxtions on Information Theory,
20 (3): 397-399, 1974.

\bibitem{Candes-and-Plan(2009)}
E. J. Candes and Y. Plan,
Near-ideal model selection by $l_1$ minimization,
The Annals of Statistics, 37: 2145-2177, 2009.


\bibitem{CaiCoherence}
T. T. Cai, G. Xu and J. Zhang,
On recovery of sparse signals via $l_1$ minimization,
IEEE Transactions on Information Theory, 55 (7): 3388-3397, 2009.


\bibitem{FoucartRIPBP}
S. Foucart, A note on guaranteed sparse recovery via $l_1$-minimization, Applied and Computational Harmonic Analysis, 29: 97-103, 2010.


\bibitem{TroppOMPCoherence}
J. A. Tropp,
Greed is good: algorithmic results for sparse approximation,
IEEE Transactions on Information Theory, 50 (10): 2231-2242, 2004.


\bibitem{WakinOMPRIP}
M. B. Wakin and M.A. Davenport,
Analysis of orthogonal matching pursuit using the restricted isometry property,
IEEE Transactions on Information Theory, 56 (9): 4395-4401, 2010.

\bibitem{FoucartRIP2010}
S. Foucart, Sparse recovery algorithms: Sufficient conditions in terms of restricted isometry constants,
in Proceedings of the 13th International Conference on Approximation Theory,
M. Neantu and L. Schumaker, eds., San Antonio, TX, 2010, Springer.

\end{thebibliography}


\clearpage

\clearpage

\begin{table} [h]
\centering
\caption{Boundedness parameters $c$ for different thersholding functions}
\label{TableBoundPar}
\vspace*{10pt}
\begin{tabular}{cccccc}
\hline
$f_{\tau,*}$   & $f_{\tau,0}$   & $f_{\tau,1/2}$     & $f_{\tau,2/3}$  & $f_{\tau,1}$  & $f_{\tau,SCAD}$   \\
\hline
$c$            &  0        & $\frac{1}{3}$      & $\frac{1}{2} $     & 1             & 1  \\
\hline
$\frac{1}{3+c}$            &  $\frac{1}{3}$        & $\frac{3}{10}$      & $\frac{2}{7} $     & $\frac{1}{4}$             & $\frac{1}{4}$   \\
\hline
\end{tabular}
\end{table}


\clearpage
\begin{center}
\text{Algorithm 1: Adaptively Iterative Thresholding Algorithm}\\ \vspace{0.2cm}
\begin{tabular}{l}
  \hline \vspace{0.2cm}
  Step 1. Normalize $A$ such that $\|A_{j}\|_{2}=1$ for $j=1, \ldots, N$; \\ \vspace{0.2cm}
  Step 2. Choose a specified sparsity level $k$ and begin with $x^{(0)}=0$;\\ \vspace{0.2cm}
  Step 3. Compute $z^{(t+1)} = x^{(t)} +  A^{T}(y - A x^{(t)})$; \\ \vspace{0.2cm}
  Step 4. Set $\tau^{(t+1)} = |z^{(t+1)}|_{[k+1]}$; \\ \vspace{0.2cm}
  Step 5. Update $x^{(t+1)} = H_{\tau^{(t+1)}}(z^{(t+1)})$; \\ \vspace{0.2cm}
  Step 6. Repeat steps 3-5 until the stop rule being satisfied; \\
  \hline
\end{tabular}\\
\end{center}

\clearpage

\begin{table} [h]
\centering
\caption{Sufficient Conditions for Different Algorithms}
\label{TableERC}
\vspace*{10pt}
\begin{tabular}{>{\scriptsize}c>{\scriptsize}c>{\scriptsize}c>{\scriptsize}c>{\scriptsize}c>{\scriptsize}c>{\scriptsize}c>{\scriptsize}c}
\hline
Algorithm   & BP    & OMP    & CoSaMP     & hard  & soft       & half   & Other AIT     \\ \hline
$\mu$      & $\frac{1}{2k^*-1}^{[28]}$    & $\frac{1}{2k^*-1}^{[32]}$  & $\frac{0.384}{4k^*-1}^\star$  & $\frac{1}{3k^*}$   & $\frac{1}{4k^*}$     & $\frac{3}{10k^*}$    & $\frac{1}{(3+c)k^*}$     \\ \hline
$(s, \delta_s)$  & $(2k^*,0.465)^{[31]}$    & $(k^*+1,\frac{1}{3\sqrt{k^*}})^{[33]}$  & $(4k^*,0.384)^{[34]}$ & $(3k^*,0.5)^{[34]}$   & -- & --  &-- \\ \hline
\end{tabular}
\vspace*{-20pt}
\end{table}
$\star$: a coherence based sufficient condition for CoSaMP derived directly by the fact that $\delta_{4k^*}<0.384$ and $\delta_s\leq (s-1)\mu$;
--: represents no related theoretical result as far as we know.

\end{document}